\magnification=\magstep1
\hsize=16 true cm
\vsize=24 true cm
\hoffset=.0 true cm        
\voffset=.0 true cm
\baselineskip=18pt
\nopagenumbers
\null
\font\tfont=cmbx12  scaled 1300           
\font\eightrm=cmr8
\font\eightbf=cmbx8
\def\title#1{ \centerline{\tfont #1}}
\def\author#1{\vskip.6cm \centerline{#1} }
\def\institution#1{\centerline{\it #1} }
\def\abstract#1{  {\vskip.6cm \noindent \baselineskip=15pt
                   \narrower {\eightbf Abstract.}{\eightrm #1}  
                   \vskip.6cm } \baselineskip=18pt    }
\def\references{\vskip.6cm \noindent{\bf References} \vskip.6cm }
\def\subsub#1{\vskip8pt{\it #1}}
\def\ibid{{\it ibid\/.~}}

\title{Dirac-like Affine Fields in 3D}
\author{I. Mi\v skovi\'c and Dj. \v Sija\v cki}
\institution{Institute of Physics, P.O.Box 57, 11001 Belgrade, Yugoslavia}

\abstract{
A generalization of the Dirac field equation in three-dimensional Minkowski
space-time to the case of the $\overline{SL}(3,R)$ $\subset$ 
$\overline{SA}(3,R)$ symmetry is considered. Constraints that ensure a correct
physical interpretation of the corresponding particle states are presented. 
Dirac-like equations based on both multiplicity-free and generic 
infinite-component $\overline{SL}(3,R)$  representations are outlined.}

{\bf Introduction.} 
The Dirac equation and the corresponding fields, that describe relativistic
point-like quantum objects of spin $1\over 2$, played a crucial role in
various important Particle Physics developments. There is a long list
starting with successful description of the electron field and its
electromagnetic interactions that goes over many aspects of the gauge
theories and the Standard Model of electroweak and strong interactions to
super-symmetry and non-commutative geometry.
The aim of this paper is to consider a generalization of the Dirac equation
and the corresponding fields in $3$-dimensional Minkowski space-time to the
case of the affine symmetry $SA(3,R)$. In this case the Lorentz subgroup
$SO(2,1)$ of the Poincar\'e group in $3$-dimensions is enlarged to the group
of all (special) linear transformations $SL(3,R)$. The relevance of such a
Dirac-like Affine Symmetry equation are for an effective description of the
IR (confining) region of the Hadronic Physics as well as in the field of
gravitational interactions of spinorial matter.

\subsub{Hadronic Matter in the IR Region.} 
There are two recent results that focus on the relevance of the group of
special linear transformations for an effective description of the hadronic
matter in QCD. (i) It has been shown, that a Yang-Mills theory based on the
$SU(2)$ group in $3D$ can be recast in the form of the General Relativity
theory [1]. In particular one finds [2] a spatial $SL(3,R)$ symmetry group.
(ii) It has been demonstrated that the QCD theory in the IR region can be
described by an effective gravity-like theory (Chromogravity) [3]. Here,
instead of the local $SU(3)$ color symmetry one has an induced $Diff(4,R)$
group of General Coordinate Transformations ($GCT$). The infinite-component
$\overline{SL}(4,R)$ $\subset$ $Diff(4,R)$ representations describe the 
hadronic matter fields [4]. 

\subsub{General Covariance and Spinorial Matter.} 
In the standard approach to General Relativity one starts with the group of
General Coordinate Transformations and the theory is set upon the principle
of general covariance. A unified description of both tensors and spinors
would require the existence of respectively tensorial and (double valued)
spinorial representations of the $GCT$ group. It is well known that the
finite-dimensional representations of $\overline{GCT}$ are characterized by
the corresponding ones of the $\overline{SL}(4,R)$ group, and
$\overline{SL}(4,R)$ does not have {\it finite} spinorial representations.
However there are infinite-dimensional spinors of $\overline{SL}(4,R)$ which
are the true "world" (holonomic) spinors [5]. The anholonomic
$\overline{SL}(4,R)$ spinors describe the fermionic matter fields of the
Metric-Affine Theory of Gravity [6].

{\bf Physical Requirements.} 
The affine group $\overline{SA}(3,R) = T_{3} \wedge \overline{SL}(3,R)$, is
a semidirect product of translations and $\overline{SL}(3,R)$ generated by 
$Q_{\mu\nu}$ ($\mu ,\nu =0, 1, 2$). The antisymmetric operators $M_{\mu\nu} = 
{1\over 2} (Q_{\mu\nu} - Q_{\nu\mu})$ generate the Lorentz subgroup 
$\overline{SO}(2,1)$, the symmetric traceless operators (shears) $T_{\mu\nu} 
= {1\over 2} (Q_{\mu\nu} + Q_{\mu\nu}) - {1\over 3} \eta_{\mu\nu} 
Q_{\sigma}^{\ \sigma}$ generate the proper $3$-volume-preserving
deformations.

\subsub{Unitarity.} 
As in the Poincar\'e case, the $\overline{SA}(3,R)$ unirreps are induced
from the unirreps of the corresponding little group $T^{\prime}_{2} \wedge 
\overline{SL}(2,R)$. In the physically most interesting case $T^{\prime}_{2}$ 
is represented trivially. The corresponding particle states have to be
described by the unitary $\overline{SL}(2,R)$ representations, which are
infinite-dimensional owing to the $\overline{SL}(2,R)$ noncompactness. 
Therefore, the corresponding $\overline{SL}(3,R)$ matter fields $\Psi (x)$ 
are necessarily infinite-dimensional and when reduced with respect to the
$\overline{SL}(2,R)$ subgroup should transform with respect to its unirreps.

\subsub{Particle Properties.}
Had the whole $\overline{SL}(3,R)$ been represented unitarily, the Lorentz
boost generators would have a hermitian intrinsic part; as a result, when
boosting a particle, one would obtain a particle with a different spin, i.e.
another particle - contrary to experience.  There exists however a
remarkable inner {\it deunitarizing} automorphism ${\cal A}$ [4], which leaves
the $R_{+} \otimes \overline{SL}(2,R)$ subgroup intact, and which maps the
$T_{0k}$, $M_{0k}$ generators into $iM_{0k}$, $iT_{0k}$ respectively 
($k=1,2$). The deunitarizing automorphism allows us to start with the
unitary representations of the $\overline{SL}(3,R)$ group, and upon its
application, to identify the finite (unitary) representations of the abstract
$\overline{SO}(3)$ compact subgroup with nonunitary representations of the
physical Lorentz group. In this way, we avoid a disease common to most of 
infinite-component wave equations, in particular those based on groups
containing the $\overline{SL}(4,R)$ group [7].

{\bf Dirac-like Infinite-component Equation.} 
Let us consider a Dirac-like equation, 
$$
(X^{\mu} p_{\mu} - M) \Psi (x) = 0
$$
for the field $\Psi (x)$ that transforms as follows,
$$
\Psi (x) \mapsto \Psi '(x') = {\cal D}(\bar A)\Psi (A(x-a)), \quad 
(a,\bar A) \in T_3 \wedge \overline{SL}(3,R)^{\cal A}.
$$
The $X_{\mu}$ matrices generalize the Dirac $\gamma_{\mu}$ ones, act in the 
space of infinite-component spinorial fields $\Psi (x)$ and ensure the 
$\overline{SL}(3,R)^{\cal A}$ covariance ($X_{\mu} \mapsto {\cal D}(\bar A) 
X_{\mu} {\cal D}^{-1}(\bar A)$). 

All $\overline{SL}(3,R)$ unirreps are known [8], and explicitly 
given in terms of the representation labels $(\sigma , \delta )$, and the 
$\overline{SO}(3)$ subgroup representations $D^{(j)}$. An arbitrary 
$3$-vector operator ($j=1$) is given by 
$$
X_{\alpha} = 
a D^{{\cal A}\ (1)}_{\ \ 0\alpha} (g) + 
b [D^{{\cal A}\ (1)}_{\ \ 1\alpha} (g) + 
   D^{{\cal A}\ (1)}_{\ \ -1\alpha} (g)], 
\quad g \in \overline{SO}(3) ,
$$ 
where, in the spherical basis, $\alpha = 0, \pm 1$. There are two distinct 
cases corresponding to the multiplicity-free and generic 
$\overline{SL}(3,R)$ representations.

\subsub{Multiplicity-free Representations Case.} 
The $\overline{SO}(3)$ representations content $\{ j \}$ of the 
multiplicity-free $\overline{SL}(3,R)$ representation is characterized by 
the $\Delta j = 2$ condition, and thus for the representation starting with 
$j = {1\over 2}$ one has $\{ j \} = \{ {1\over 2}, {5\over 2}, {9\over 2}, 
\dots \}$. The matrix elements of the $X_{\alpha}$ vector operator read 
$$
\left< {{(\sigma ')}\atop {j' m'}}\right| 
X_{\alpha} 
\left|{{(\sigma )}\atop{j m}}\right> 
= a^{(\sigma ' \sigma )}_{j'j}(-)^{j'-m'}\sqrt{(2j'+1)(2j+1)} 
\left( \matrix{\ j' & 1 & j \cr -m' & \alpha & m \cr} \right)^{\cal A} .
$$

\subsub{Generic Representation Case.} 
In the nontrivial-multiplicity case we obtain the following expression for 
the $X_{\mu}$ matrix elements:
$$
\left< {{(\sigma '\ \delta ')}\atop{j'\ k'\ m'}} \right| X_{\alpha} \left| 
{{(\sigma\ \delta )}\atop{j\ k\ m}} \right> = 
(-)^{j'-k'} (-)^{j'-m'} \sqrt{(2j'+1)(2j+1)} 
\left( \matrix{ \ j' & 1 & j \cr -m' & \alpha & m \cr} \right)^{\cal A} \times
$$
$$
\times 
\Big\{ a^{(\sigma ' \delta ' \sigma \delta )}_{j'j} 
\left(\matrix{\ j' & 1 & j \cr -k' & 0 & k \cr}\right) 
+ b^{(\sigma ' \delta ' \sigma \delta )}_{j'j} \Big[ 
   \left(\matrix{\ j' & 1 & j \cr -k' & 1 & k \cr} \right)
   + \left(\matrix{\ j' & \ 1 & j \cr -k' & -1 & k \cr} \right) 
  \Big]\Big\}.
$$

The reduced matrix elements $a^{(\sigma '\delta '\sigma \delta )}_{j'j}$ and 
$b^{(\sigma '\delta '\sigma \delta )}_{j'j}$ are determined by first
embedding the $\overline{SL}(3,R)$ group into the $\overline{SL}(4,R)$ one,
then by identifying $X_{\mu}$ as $Q_{\mu 4}$ generators, and finally by 
reducing the $\overline{SL}(4,R)$ representations down to the 
$\overline{SL}(3,R)$ ones. 

In conclusion, we have shown explicitly the existence of a non-trivial
Dirac-like $\overline{SL}(3,R)$ covariant field equation fulfilling all
relevant physical requirements.

\references

\item{[1]} F.A. Lunev, Phys. Lett. {\bf B 295} 92 (1992); 
   D.Z. Freedman, P.E. Haagensen, K. Johnson and J.I. Latorre,
   hep-th/9309045; 
   E.W. Mielke, Y.N. Obukhov and F.W Hehl, Phys. Lett {\bf A 192} 153 (1994); 
   V. Radovanovi\'c and Dj. \v Sija\v cki, Class. Quant. Grav. {\bf 12} 1791
   (1995).

\item{[2]} M. Bauer, D.Z. Freedman and P.E. Haagensen, Nucl. Phys. 
   {\bf B 428} 147 (1994). 

\item{[3]} Dj. \v Sija\v cki and Y. Ne'eman, Phys. Lett. {\bf B 247} 571
   (1990); 
   Y. Ne'eman and Dj. \v Sija\v cki, Phys. Lett. {\bf B 276} 173 (1992); 
   \ibid Int. J. Mod. Phys. {\bf A 10} 4399 (1995); 
   \ibid Mod. Phys. Lett. {\bf A 11} 217 (1996).

\item{[4]} Y. Ne'eman and Dj. \v Sija\v cki, Phys. Rev. {\bf D 37} 3267 
   (1988); 
   \ibid {\bf D 47} 4133 (1993).

\item{[5]} Y. Ne'eman and Dj. \v Sija\v cki, Phys. Lett. {\bf B 157} 275
   (1985); \ibid Found. Phys. {\bf 27} 1105 (1997); 
   Dj. \v Sija\v cki, Acta Phys. Pol. {\bf B 29} 1089 (1998). 

\item{[6]} F.W. Hehl, G.D. Kerlick and P. von der Heyde, Phys. Lett. 
   {\bf B 63} 446 (1976); 
   Y. Ne'eman and Dj. \v Sija\v cki, Ann. Phys. (N.Y.) {\bf 120} 292 (1979); 
   F.W. Hehl, J.D. McCrea, E.W. Mielke and Y. Ne'eman, Phys. Rep. {\bf 258}
   1 (1995). 

\item{[7]} J. Mickelsson, Commun. Math. Phys. {\bf 88} 551 (1983); 
   A. Cant and Y. Ne'eman, J. Math. Phys. {\bf 26} 3180 (1985).

\item{[8]} Dj. \v Sija\v cki, J. Math. Phys. {\bf 16} 298 (1975); 
   Dj. \v Sija\v cki, J. Math. Phys. {\bf 31 } (1990) 1872. 

\end